\documentclass[10pt,conference]{IEEEtran}
\IEEEoverridecommandlockouts

\usepackage{booktabs} 
\usepackage{color}
\usepackage{listings}
\usepackage{enumitem}
\usepackage{threeparttable}
\usepackage{multirow}
\usepackage{hhline}
\usepackage{subfigure}
\usepackage{array}
\usepackage{pifont}
\usepackage{booktabs}
\usepackage{xcolor}
\usepackage{algorithm}
\usepackage{algorithmicx}
\usepackage[noend]{algpseudocode}
\usepackage{graphics}
\usepackage{epsfig}

\usepackage{amsmath}

\usepackage{hyperref}
\usepackage{bbding}
\usepackage{caption}

\newcommand{\pre}[1]{}

\newcommand{\rev}[1]{#1}

\newcommand{\CodeIn}[1]{{\small\texttt{#1}}}

\newcommand{\toolname}{Isra}
\newcommand{\Luyao}[1]{{\color{blue} \bf \{Luyao: {#1}\}}}
\newcommand{\Ziheng}[1]{{\color{cyan} \bf \{Ziheng: {#1}\}}}
\newcommand{\bugnumber}{33}
\newcommand{\unknownbug}{24}

\newcommand{\opnumber}{65}

\newcommand{\tone}{graph-level constraint resolving}
\newcommand{\ttwo}{operation-level constraint resolving}
\newcommand{\toneandtwo}{graph-level and operation-level constraint resolving}

\newcommand{\firstp}{graph-level directional consistency}

\newcommand{\bone}{declarative-style generation}
\newcommand{\boneshort}{DeclGen}
\newcommand{\btwo}{Randoop-like generation}
\newcommand{\btwoshort}{Randoop-Gen}

\definecolor{codegreen}{rgb}{0,0.6,0}
\definecolor{codegray}{rgb}{0.5,0.5,0.5}
\definecolor{codepurple}{rgb}{0.58,0,0.82}
\definecolor{backcolour}{rgb}{1,1,1}
 
\lstdefinestyle{mystyle}{
    backgroundcolor=\color{backcolour},   
    commentstyle=\color{codegreen},
    keywordstyle=\color{magenta},
    numberstyle=\tiny\color{codegray},
    stringstyle=\color{codepurple},
    basicstyle=\ttfamily\footnotesize,
    breakatwhitespace=false,         
    breaklines=true,                 
    captionpos=b,                    
    keepspaces=true,                 
    numbers=left,                    
    numbersep=5pt,                  
    showspaces=false,                
    showstringspaces=false,
    showtabs=false,                  
    tabsize=2
}
 
\def\BibTeX{{\rm B\kern-.05em{\sc i\kern-.025em b}\kern-.08em
    T\kern-.1667em\lower.7ex\hbox{E}\kern-.125emX}}
\begin{document}

\title{Effective Random Test Generation for Deep Learning Compilers}

\author{\IEEEauthorblockN{Luyao Ren, Ziheng Wang, Yingfei Xiong}
\IEEEauthorblockA{
\textit{Peking University}}
\and
\IEEEauthorblockN{Li Zhang, Guoyue Jiang}
\IEEEauthorblockA{
\textit{Sophgo Technologies Ltd}}
\and
\IEEEauthorblockN{Tao Xie}
\IEEEauthorblockA{
\textit{Peking University}}
}

\maketitle

\begin{abstract}
  Deep learning compilers help address the difficulties of deploying deep learning models on diverse types of hardware. Testing deep learning compilers is highly crucial, because they are impacting countless AI applications that use them for model optimization and deployment. To test deep learning compilers, random testing, the testing method popularly used for compiler testing practices, faces the challenge of generating semantically valid test inputs, i.e., deep learning models that satisfy the semantic model specifications (in short as semantic specifications). To tackle this challenge, in this paper, we propose a novel approach named \toolname{}, including a domain-specific constraint solver that resolves the constraints from the semantic specifications without backtracking. We implement and apply our approach to three popular real-world deep learning compilers including TVM, Glow, and a commercial compiler named SophGo.
  The evaluation results show that \toolname{} is more effective than the state-of-the-art approaches and the baseline approaches on constructing valid test inputs for compiler-bug detection, and \toolname{} successfully finds \unknownbug{} previously unknown bugs in released versions of the three compilers. These results indicate \toolname{}'s effectiveness and practical value.
\end{abstract}


\section{Introduction}
In recent years, deep learning has been widely used in software systems from various domains, such as autonomous driving, e-commerce, and smart cities.
Given that deep learning models become increasingly large and complicated, there are emerging needs to deploy versatile deep learning models on different types of hardware such as GPU, FPGA, and TPU~\cite{DBLP:conf/isca/JouppiYPPABBBBB17}.
To reduce the burden of optimizing deep learning models and to address the difficulty of model deployment on hardware, deep learning compilers have been developed, such as TVM~\cite{DBLP:conf/osdi/ChenMJZYSCWHCGK18}, Glow~\cite{DBLP:journals/corr/abs-1805-00907},
and XLA~\cite{xla}.
These deep learning compilers have been widely used for the optimization and deployment of deep learning models, especially those with critical performance requirements.


Testing a deep learning compiler is vital for two main reasons. First, if a deep learning compiler used by AI applications contains bugs, the deployed AI applications can exhibit serious failing behaviors. For example, a critical bug of TVM's SPIRV codegen led to incorrect results for a TVM-optimized model's output, which affected all users who use TVM for their model deployment on the Nvidia Vulkan backend\footnote{\url{https://github.com/apache/tvm/pull/8102}}. Second, the highly sophisticated compilation process in a deep learning compiler is error-prone. According to a very recent study~\cite{DBLP:conf/sigsoft/ShenM0TCC21}, during a time period of 15 months, there are 845 bug-fixing pull requests on the TVM project, including 318 bugs in total.

A deep learning model, as the input of a deep learning compiler, needs to satisfy semantic specifications in two aspects; otherwise, it will be rejected by the deep learning compilers at an early stage before invoking the actual core functionality of the compilers.
First, a deep learning model is a neural network arranged as a directed and acyclic graph. Second, the model also needs to satisfy certain constraints, which are required by the operations in the model. For example, within a model, a \CodeIn{MatMul} operation (denoting matrix multiplication) with two input matrices (2-D tensors) requires that the number of columns in the first matrix is equal to the number of rows in the second matrix.

To test core compiler parts (achievable by only valid inputs),  
one can indeed adopt random testing (the most popularly used technology for compiler testing practices~\cite{DBLP:journals/csur/ChenPPXZHZ20}) in a straightforward way: generating possible inputs and filtering them by checking against the semantic specifications\footnote{Checking against the semantic specifications can be in the form of a boolean checker such as \CodeIn{repOK}~\cite{boyapati02:korat}, etc.}, also called as declarative-style random test generation~\cite{boyapati02:korat, DBLP:conf/issta/ElkarabliehMK08, DBLP:conf/icst/GligoricGLMK09}; however, this style suffers from two main issues.
First, random testing in the declarative style has a fairly low probability of satisfying the semantic specifications, especially for complicated operations (as shown in our experimental results in Section~\ref{sec:evaluation}), wasting testing the budget on generating and checking many invalid inputs.
Second, valid inputs generated by this random testing strategy tend to be simple, as complex inputs have an even lower probability of satisfying the semantic specifications, whereas it is highly critical to also generate complicated models in order to achieve various testing goals.

To better conduct random testing on a deep learning compiler, we take the semantic specifications of a deep learning model as logic constraints; in this way, test generation is equivalent to finding solutions to the constraints. However, we face a challenge due to complex constraints, i.e., those related to both the graph structure of the model and operations within the model. Furthermore, within the constraints, the involved first-order/second-order logic (as shown in Section~\ref{sec:s2}) is  undecidable in general~\cite{turing1936computable} and causes existing solvers not to be able to encode, or perform efficiently ~\cite{DBLP:conf/iccad/DutraBS18, gradel2007finite}.

To address the preceding challenge, we propose a novel  approach named \toolname{} based on the following insight: 
the constraints on a deep learning model have certain properties, allowing us to iteratively resolve and simplify the constraints to effectively find solutions, by following a proper instantiation order.
We design two strategies in the core part of \toolname{}, a novel domain-specific constraint solver. Our solver conducts instantiation with an order for gradually resolving and simplifying constraints.
Based on the consistency among the constraints, \toolname{}, with our domain-specific constraint solver, is able to find semantically valid inputs without backtracking, while  ensuring both soundness (the generated inputs are semantically valid) and completeness (no loss for the probability of generating any valid model).

To evaluate \toolname{},  we implement it and empirically compare it with \pre{four} \rev{five} baselines: (1) the aforementioned approach of random test generation, named as \bone, (2) test generation based on the idea of feedback-guided test generation~\cite{pacheco2007feedback}, named \btwo, (3)  a  state-of-the-art tool named  Muffin~\cite{DBLP:conf/icse/GuLZ022} implementing a generation-based approach for deep learning models, and (4) \rev{a mutation-based approach,} TVMFuzz~\cite{DBLP:conf/sigsoft/ShenM0TCC21}, \pre{the only existing tool} toward testing deep learning compilers \rev{, (5) a state-of-the-art generation-based approach named NNSmith~\cite{NNSmith} toward testing deep learning compilers}.
Our evaluation results show that our \toolname{} approach substantially outperforms the baselines on the metrics of generated valid test inputs under the same setting, for demonstrating our approach's effectiveness.
Furthermore, to investigate the bug detection capability, when used to test the same benchmark (TVM, Glow, and SophGo), \toolname{} detects \bugnumber{} unique bugs in total (with 18 on TVM, 4 on Glow, and 11 on SophGo), \pre{outperforming} \rev{performing better or as well than} the baselines.

In addition, among the bugs found by \toolname{}, there are \unknownbug{} previously unknown bugs. After these previously unknown bugs were reported to compiler developers, 19 were confirmed and 16 were already fixed upon our bug reporting so far. 
The positive feedback from the developers also shows \toolname{}'s high value in practice. The source code, experimental results, and bug reports are publicly available at \url{https://github.com/israProj/isra}.

In summary, this paper makes the following contributions:
\begin{itemize}
	\item An effective test generation approach named \toolname{} for testing deep learning compilers, based on instantiation-based constraint solving, working in a backtrack-free way, with the guarantee of soundness and completeness.
	\item A domain-specific constraint solver toward  finding solutions to the constraints specified from the semantic specifications of a deep learning model, with two novel strategies for resolving complex constraints.
	\item Implementation and evaluation of our approach for showing its high effectiveness and high practical value, including
 \pre{outperforming state-of-the-art approaches (Muffin~\cite{DBLP:conf/icse/GuLZ022} and 
 TVMFuzz~\cite{DBLP:conf/sigsoft/ShenM0TCC21})}
 \rev{outperforming state-of-the-art approaches on coverage metrics, achieving comparable and complementary results on the bug detection}, and successfully finding \bugnumber{} unique bugs in three popular real-world deep learning compilers.
\end{itemize}


\section{Background and Overview}\label{sec:s2}

\begin{figure}[t]
    \centering
    \includegraphics[width=\linewidth]{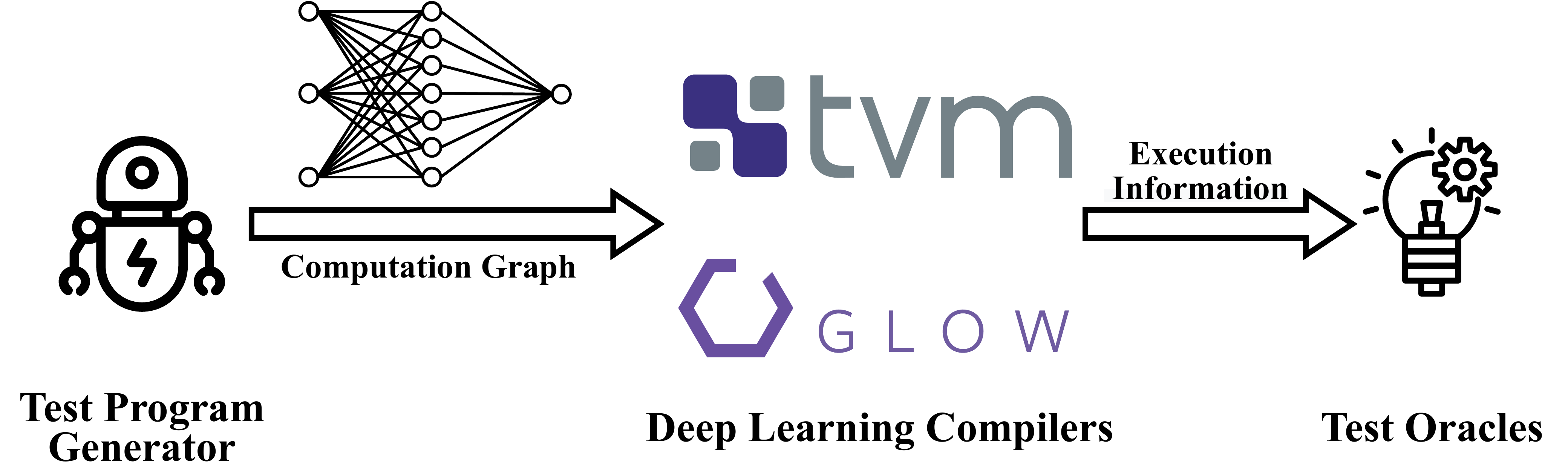}
    \caption{The Pipeline of Testing Deep Learning Compilers}
    \label{fig:pipeline}
    \vspace{-0.3cm}
\end{figure}

Figure~\ref{fig:pipeline} shows our overall pipeline of random testing for deep learning compilers. In the stage of test generation, we use our test program generator to generate random computation graphs that are semantically valid. For the test oracle, we use both differential testing and the crash status of the compiler under test~\cite{DBLP:journals/csur/ChenPPXZHZ20}.  
Before formally describing our approach in detail (as shown in Section ~\ref{sec:approach}), we first take an overview of background with specific examples, then illustrate our approach in a nutshell.

\subsection{Computational Graph for Deep Learning}\label{sec:s2.1}

A deep learning model, as the input of deep learning compiler, can be regarded as a computation graph, where an edge represents a tensor (an N-D array), denoting the flow of data between nodes, and a node represents (1) an operation (denotes a mathematical function) whose inputs are the incoming edge(s) and outputs are the outcoming edge(s), or (2) a tensor placeholder where we need to represent creation of a user-defined variable such as the input tensor of computational graph.
The computation graph can be executed by feeding the specific data into tensor placeholders. 
The formal definitions of the computation graph are shown in Section~\ref{sec:approach}. 

\begin{figure}[t]
    \vspace{-0.2cm} 
    \centering
    \setlength{\abovecaptionskip}{0cm}   
    \setlength{\belowdisplayskip}{0pt} 
    \includegraphics[width=\linewidth]{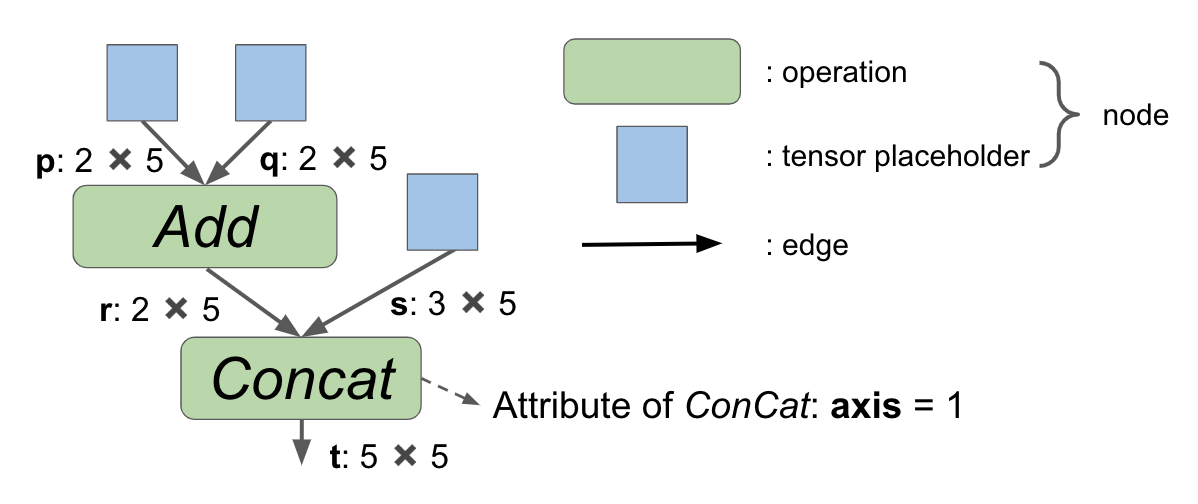}
    \caption{An Example of a Computation Graph.}
    \label{fig:p2}
    \vspace{-0.3cm}
\end{figure}

As an example, Figure~\ref{fig:p2} shows a deep learning model with two operations. The first operation is \CodeIn{Add}. It takes two tensors $p$ and $q$ as its input and outputs a tensor $r$ as their sum. The second operation is \CodeIn{Concat}. It accompanies an attribute $axis$ (denoting the dimension of the axis to concatenate on) and takes two tensors $r$ and $s$ as its input, and outputs a tensor $t$ as their concatenated results. The edge in the computation graph represents the dataflow that gets transferred between the nodes. For example, $r$, as the output of \CodeIn{Add} operation, could be transferred to the input of \CodeIn{Concat} operation.


A computation graph is directed and acyclic, specifying the order of computation.
In the example, you need to compute \CodeIn{Add} first in order to compute \CodeIn{Concat} because the output of \CodeIn{Add} (i.e., tensor $r$) flows to the input of \CodeIn{Concat}.
Except for the acyclicity of the graph, for each operation, the number of incoming edges should be aligned with the number of input arguments defined by corresponding mathematical function that operation denotes. For example, \CodeIn{Concat} requires two or more input arguments, so the number of incoming edges should be more than or equal to two. We called those semantic specification of computation graphs as \textbf{graph-level constraints}.

Besides graph-level constraints, each operations in the computation graph holds its internal semantic specification, specified from the definition of mathematical function that operation denotes, which we called \textbf{operation-level constraints}.
In our example, as the input of \CodeIn{Add} operation, tensor $p$ and $q$ should have the same shape.
Similarly, as for \CodeIn{Concat} operation, tensor $r$ and $s$ must have the same shape, except for the dimension of the axis to concatenate on (defined by an operation's attribute $axis$).
Particularly, by explicitly denoting the structure of tensors, operation-level constraints of the computation graph in Figure~\ref{fig:p2} could be specified as follows ($dim_a$ denotes the dimension size of tensor $a$, and $a[i]$ denotes the length of the i-th dimension of tensor $a$) :
\begin{eqnarray}
& dim_{p} = dim_{q} = dim_{r} \label{e3} \\
& \forall i \in [1, 2, ..., dim_p], p[i] = q[i] = r[i] \label{e4}  \\
& dim_{r} = dim_{s} = dim_{t} \label{e5} \\
& \forall i \in [1, 2, ..., dim_r] \wedge i \neq axis, r[i] = s[i] = t[i] \label{e6} \\
& t[axis] = r[axis] + s[axis] \label{e7}\\
& 1 \leq axis \leq dim_r \label{e8}
\end{eqnarray}

\subsection{Challenges}\label{sec:s2.2}


The complicated semantic specifications of the computation graph, which consist of both the graph-level constraints and operation-level constraints, result in the sparsity of semantically valid inputs (compared to syntactically valid inputs).
Thus, random test generation suffers the issues on effectiveness. Specifically, the approach that randomly generates possible computation graphs, and then filters invalid ones by checking against the semantic specification, holds fairly low possibility to satisfy the semantic specification.
As our previous example, for two tensors of the input of an \CodeIn{Add} operation, assume that the range of the tensor's dimension size is $O(D)$ and the length of each dimension is $O(L)$, the generation holds $O(L^{-D})$ possibility of producing valid ones due to Equation~\ref{e4}.
The possibility diminishes with larger deep learning model, wider range, or more complex specifications.


In order to better conduct random testing on deep learning compilers, instead of taking semantic specification as a blackbox checker, we explicitly specify the semantic specifications of computation graph as constraints, i.e., take semantic specification as an whitebox, and test generation is equivalent to find solutions that satisfies those constraints.

However, existing practices of constraint solving are limited due to our complex constraints which are expressed in first-order/second-order logic instead of propositional logic due to following reasons: (1) the acyclicity of computation graph;
(2) the existence of quantifiers such as $\forall i \in [1, 2, ..., dim_p]$ in Equation~\ref{e4}; (3) the existence of unknown functions such as $r[axis]$ in Equation~\ref{e7} (it is called unknown function because we actually need to construct a function that maps to the length of each dimension of a tensor that we may not know the dimension size).
Compared with propositional logic that is decidable, solving first-order/second-order logic (with quantifiers and unknown functions) is challenging because theoretically the first-order/second-order logic is undecidable in general~\cite{turing1936computable} and also, in practice, quantifiers and unknown functions cause existing solvers unable to encode, or perform inefficiently ~\cite{gradel2007finite, DBLP:conf/cade/ReynoldsTGKDB13, DBLP:conf/iccad/DutraBS18}.

\subsection{Instantiation-based Constraint Solving} \label{sec:instantiation}
Instantiation ~\cite{apt2003principles, kumar1992algorithms, prosser1993hybrid, reynolds2014finding, reynolds2013quantifier} is a widely used technique for solving constraint satisfaction problem (CSP) such as satisfiability modulo theories (SMT).
By assigning the values to the variables in the constraints, we could get an instantiation of the constraints.

An instantiation-based solver starts from an empty instantiation and extends instantiation (mostly in an iterative way) to find solutions. 
An instantiation could be extended by assigning the values to the variables that are not assigned in the instantiation before.
In the meanwhile, by replacing the variables with their assigned specific values in the constraints, also, with the help of solver, it is possible to simplify constraints and reduce the domain of unassigned variables (it is called as constraint propagation~\cite{apt2003principles}). 
For example, for the constraint $S \subset T$ ($S$ and $T$ are set variables), if we first instantiate $T$ as $\{1,2\}$, then it is easy to simplify the constraints by solving constraints and simultaneously deducing the domain of $S$, i.e., $S \subset T \Rightarrow S \subset \{1,2\} \Rightarrow S \in \{\emptyset,\{1\},\{2\}\}$.

With instantiation, the elimination technique has a chance to be applied for the simplification on constraints which are originally encoded in first-order logic (or higher-order logic). Inspired by quantifier elimination~\cite{DBLP:conf/cade/MouraB07, gradel2007finite}, if we already determine the domains of quantifiers or unknown functions in constraints according to the instantiation, we could then simplify the constraint by rewriting the constraints to an quantifier-free and unknown-function-free form. 
For example, assume $S$ is a set variable, for the constraint $\forall x \in S, P(x)$, if we already instantiate the domain of the universal quantifier $x$ as $S: \{1,2,3\}$, then we could eliminate the quantifier by rewriting the constraints as follows: $\forall x \in S, P(x) \Longrightarrow P(1) \wedge P(2) \wedge P(3)$. 

We call an instantiation consistent if it could be extended to a solution, otherwise it is inconsistent. For example, in the constraint $S \subset T$, if we first instantiate $T$ as $\emptyset$, then the instantiation is inconsistent.
Generally, instantiation-based solvers may backtrack to try other instantiations if it finds instantiations are inconsistent. Backtracking decreases the solver's efficiency on finding solutions.

\subsection{\toolname{} in a Nutshell}\label{sec:nutshell}


To effectively generate semantically valid computation graphs, we propose an effective random test generation approach, named \toolname{}, including a domain-specific constraint solver with two strategies: \toneandtwo{}, based on our key idea that constraints are able to be simplified with a well-designed instantiation order. Next, we introduce a running example to briefly illustrate how \toolname{} works.




\begin{figure*}[htp]
    \centering
    \setlength{\abovecaptionskip}{0.2cm}   
    \setlength{\belowdisplayskip}{0pt} 
    \includegraphics[width=\linewidth]{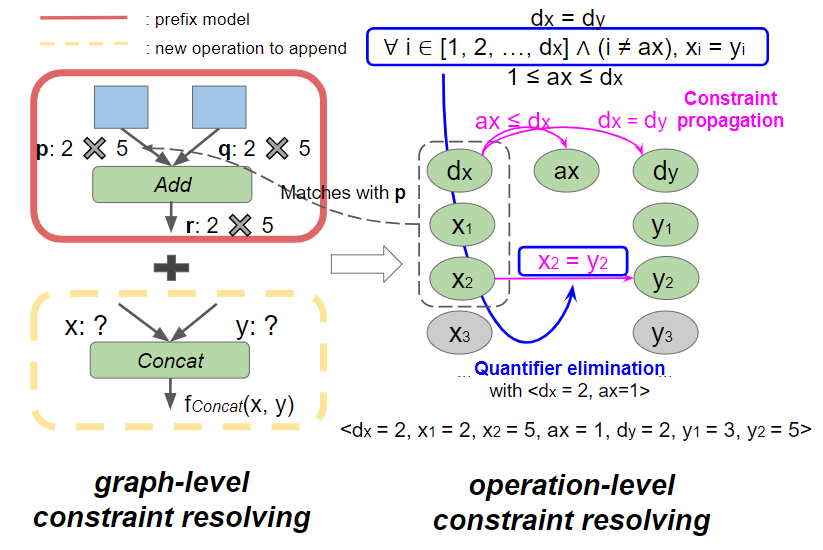}
    \vspace{-0.3cm} 
    \caption{A running example of \toolname{}}
    \label{fig:p3}
\end{figure*}


\subsubsection{Graph-level constraint resolving}

Our generation follows a topological order of operations in computation graph. We say node a precedes node b, or b succeeds a, if there exists a path, where a appears earlier than b in the path. Each time we generate a node, this node does not precedes any existing node. 
For example, as the graph shown in Figure~\ref{fig:p2}, followed by a topological order, i.e., operation $Add$ then operation $Concat$, our approach instantiates operations one by one. For each operations, we first instantiate its type and the number of incoming edges.

In this way, our approach resolves the constraints by partitioning them into several subparts. Each subpart corresponds to a single operation and its related edges. Furthermore, because the output of operations could be determined only by the input and attributes, we rewrite the constraints by substituting the output as a function of the input and attributes. In our example, the constraints are as follows, where $Spec_{op}(V)$ is defined as a set of constraints on $V$ that specifies from the specification of an operation with type $op$, $f_{op}$ denotes the mathematical function of the operation with type $op$:

\begin{eqnarray}
& S_1: Spec_{Add}(p,q) \wedge (r = f_{Add}(p,q)) \label{eqs1} \\
& S_2: Spec_{Concat}(axis,r,s) \wedge (t = f_{Concat}(axis, r, s)) \label{eqs2}
\end{eqnarray}


After resolving constraints with instantiating operations in the graph by their topological order, our goal turns to instantiate each single operation by solving constraints related to the operation (in our example, they are $Spec_{Add}(p,q)$ and $Spec_{Concat}(axis,r,s)$), as shown in the next part.

\subsubsection{Operation-level constraint resolving}
The instantiation of a new operation includes assigning the value to its operation type, attributes, input (incoming edge(s)) (output of the operation is excluded as explained before).
As a concrete example, assume we already instantiate a computation graph with a single \CodeIn{Add} operation (with $p$ and $q$ as its input as shown in the red part of Figure~\ref{fig:p3}), now we extend the instantiation by appending a new operation into the computation graph.


Assume we instantiate type of the new operation as \CodeIn{Concat}, and the number of incoming edges of the new operation as two in our example.
We now set symbol variables for the input and attributes of the operation.
In the example of \CodeIn{Concat}, we denote variable $ax$ for the attribute $axis$, and variable $x$ and $y$ as two incoming edges (note, $ax$, $x$ and $y$ are just symbol variables, which will be assigned with values later, such as assigning $r$ to $x$).

For each incoming edge, i.e., each tensor in the input, instead of instantiating the whole tensor, we focus on instantiating the structure of the tensor first due to that the semantic specifications are only related to tensor structure. For an edge $x$, we setup a set of symbol variables to substitute $x$ in the constraints, including $d_x$ (denoting the dimension size of $x$, i.e., $dim_x$) and $x_i$ (denoting each dimension's length of $x$, i.e., $x[i]$). In this way, the constraints could be further specified as follows: 
\begin{eqnarray}
& \bigwedge_{k \in \{1, 2, 3\}} C_k(ax, d_x, d_y, x_i, y_i)\\
& C_1: d_{x} = d_{y} \label{c1} \\
& C_2: \forall i \in [1, 2, ..., d_x] \wedge (i \neq ax), x_i = y_i \label{c2} \\
& C_3: 1 \leq ax \leq d_x \label{c3}
\end{eqnarray}

To sample a random solution to the above constraints, our approach instantiates variables in a well-designed order:
first are variables related to $x$; then attribute $axis$; finally variables related to $y$. In our example, the order is as follows: $d_x$; $x_i$; $axis$; $d_y$; $y_i$.
Note, in the order, variables related to the same tensor are ordered together in a group, also, within the group, instantiation of the dimension size (e.g., $d_x$) is ahead of the length of each dimension (e.g., $x_i$). The detailed illustration and explanation of the order are shown in Section~\ref{sec:s3.3}.

Followed by this order, we are able to simplify the constraints to quantifier-free and unknown-function-free by the elimination technique mentioned in Section~\ref{sec:instantiation}; also controllably choose ways for instantiating unassigned tensors, i.e., instantiating the tensor as an instantiated one such as $r$ for $x$; or as the output from a tensor placeholder such as $s$ for $y$. Details are shown in Section~\ref{sec:approach}.

With above simplification, the constraints belong to propagation logic which is decidable. Thus, we are able to conduct constraint solving by constraint propagation~\cite{apt2003principles} to find solutions. In addition, the constraint propagation will not produce an empty domain (which causes the instantiation inconsistent) due to the good property as explained in next part, resulting in the overall process being backtrack-free.

\subsubsection{Properties of \toolname{}}\label{sec:properties}
Based on the graph theory and the theory of constraint satisfaction problem (CSP)~\cite{apt2003principles}, our instantiation-based solver holds some good properties due to characteristics of constraints on deep learning models. We draw main conclusions here, formal definitions and detailed explanations are shown in Section~\ref{sec:approach}.

The first property is called \firstp{}. For any semantically valid computation graph with a topological order on operations as  $(O_1, O_2, ..., O_i)$, we can consistently extend the instantiation to include $O_{i+1}$ (with topological order as $(O_1, O_2, ..., O_i, O_{i+1})$) as long as ensuring satisfaction of constraints related to $O_{i+1}$.

The second property is called operation-level global consistency. For constraints related to each single operation, after determining operation's type and the number of tensors in the input, by taking attributes as a variable and each tensor in the input of the operation as a variable, this CSP, which consists of constraints on the operation level, is globally consistent: any consistent instantiation of a subset of the variables can be extended to a consistent instantiation of all of the variables without backtracking~\cite{dechter1992local}.

\section{Detailed Approach Description}\label{sec:approach}




\subsection{Notations and Definitions}\label{sec:s3.1}

\subsubsection{Concepts in the Computation Graph}

A tensor $t$ is a generalized vector, like N-D array (N is a positive integer). The structure of tensor $t$ is defined as a set $Str_t$, denoting the structural information of the tensor. $Str_t$ includes (1) a numerical value $dim_t$ that denotes $t$'s dimension size and (2) a variable-length array that denotes each dimension's length of tensor $t$ (i.e., $t[i]$ as the i-st dimension's length).  

An operation $n$ is defined as a function with tensor(s) as input and output. It has some parameters: we denote $Op_{n}$ as its type ($Op_n \in AllOps$, $AllOps$ is a unverisal set which contains all types of operations), $Attr_{n}$ as a set which contains attributes of the operation (such as \emph{strides} in \CodeIn{Conv} operation, and \emph{axis} in \CodeIn{SoftMax} operation). 
Also, $Input_n=\{in_1, in_2, ...\}$ is a set which contains one or more tensors as the input of $n$, and $Output_n=\{out_1, out_2, ...\}$ as the output of $n$. 
Specially, $Attr_n$ contains a special attribute $Indegree_n$ as the number of tensors in $Input_n$, i.e., $Indegree_n = |Input_n|$. 

A tensor placeholder $tp$ is simply a variable, denoting a tensor to which the specific data will be assigned later. A tensor placeholder that denotes tensor $p$ could be created with merely $Str_p$, without need of specific data. 


A computation graph $G$ is defined as an ordered triple $(V_G, E_G, \psi_G)$, where the set $V_G$ denotes the nodes, the set $E_G$ denotes the edges. An element in $V_G$ is either an operation or a tensor placeholder. An element in $E_G$ is a tensor. $\psi_G$ is called an incidence function which maps an edge into a pair of nodes (i.e., a mapping from $E(G) \rightarrow V(G) \times V(G)$), denoting the structure of the graph.



\subsubsection{Constraint Satisfaction Problem}

A constraint $C$ is a limitation placed on the values of variables. Consider a finite sequence of variables $S:=\{x_1,x_2,...,x_k\}$, with respective domains $D(x_1)$, . . ., $D(x_k)$ associated with them. So each variable $x_i$ ranges over the domain $D(x_i)$. By a constraint $C$ on $S$ we mean a subset of $D(x_1) \times D(x_2)... \times D(x_k)$.


An instantiation $Q$ is defined as a set of tuples, written as $\left \langle x_1 = v_1, x_2 = v_2, ..., x_m = v_m \right \rangle$, denoting that a specific value $v_i$ that has been assigned to the variable $x_i$. 

A constraint satisfaction problem (CSP) $P:(V, D, C)$, where $V$ is a set of variables, $D$ is the set of domains of values for each variable, $C$ is a set of constraints.
A solution to a problem $P$ is an instantiation $Q$, containing the assignment of all variables in $V$, which satisfies all of constraints in $SC$. Namely, an instantiation $Q$ is a solution of $P: (V, D, C)$ only if it satisfies with all of constraints in $C$.

\subsubsection{Local Consistency and Global Consistency}
Let $X = (X_1, X_2, ..., X_n) $ be a set of variables in a CSP, and let $X^{'} = (X^{'}_1, X^{'}_2, ..., X^{'}_m)$ be a subset of variables from $X$.  A partial instantiation of variables $\left \langle X^{'}_1 = x_1, X^{'}_2 = x_2, ..., X^{'}_m = x_m \right \rangle$ is locally consistent if it satisfies all the constraints in $X^{'}$.
A globally consistent CSP is one in which any locally consistent partial instantiation of variables can be extended to a consistent full instantiation. Globally consistent CSPs have the property that a solution can be found without backtracking ~\cite{dechter1992local}.

\subsection{Graph-level constraint resolving}\label{sec:s3.2}

Based on the acyclic trait of the computation graph, for any computation graph, there always exists a topological order of operations, i.e., for every two operation $x$ and $y$ in computation graph, if there is a tensor that is both the output tensor of $x$ and the input of $y$, then operation $x$ comes before operation $y$ in the ordering.

Our approach works as a top-down way to incrementally instantiate a computation graph by iteratively instantiate a new operation and appending it into the computation graph, as shown in Figure~\ref{fig:p3}. 
Specifically, we follow the topological order of operations to generate them in the computation graph. When to generate a new operation $x$, we need to instantiate $OP_x$, $Attr_x$ and $Input_x$, with ensuring the satisfaction $Spec_{OP_x}(Attr_x, Input_x)$ (the same with the definition in Section~\ref{sec:nutshell}). We leave the instantiation of edges in $Output_x$ later (when we instantiate another operation $y$ with an edge from $x$ to $y$) because the value of $Output_x$ could be determined only by $OP_x$, $Attr_x$ and $Input_x$, i.e., $Output_x = f_{OP_x}(Attr_x, Input_x)$.



After finishing the instantiation for current the operation (with its attributes and incoming edges), we iterate the same process for instantiating the next operation if the number of operations in the computation graph has not exceeded to the parameter we set (named $numop_{G}$ as the number of operations in generated computation graph).

\subsubsection{Property of \firstp{}}


Taking each operation as a variable (a set variable $V_x$, containing $Op_x$, $Input_x$, $Attr_x$), constraints of the computation graph could be rewritten as a binary CSP, consisting of two kinds of constraints: first are unary constraints on each variable, and second are binary constraints between two variables (i.e., the output of an operation equals to the input of another operation).
According to the definition, an instantiation $Q$ is locally consistent on $V_x$ if $Q$ satisfies all the constraints in $V_x$.

Because our instantiation follows the topological order of a computation graph, with only instantiating edges which are from previous nodes (variables that are ahead of $V_x$ in the order) to current variable $V_x$, the instantiation on $V_x$ will not affect local consistency of previous variables (based on the property of topological order). Thus, for each $x$, as long as we are able to instantiate $Q$ with its local consistency on $V_x$ every time, then we could include $V_x$ in the end of topological order, and instantiation is still consistent. Also, because we instantiate edges with the direction followed by the topological order, the overall graph is loop-free. Thus, following with the topological order, ensuring local consistency on each operation leads to bracktrack-free instantiation on the graph level.

\subsection{Operation-level constraint resolving}\label{sec:s3.3}

To instantiate a new operation $x$ and append it in the existing computation graph, we need to instantiate (1) $OP_x$, (2) $Attr_x$, (3) $Input_x$ (i.e., incoming edges). We take the instantiation of these items into following steps.

We first determine the $OP_x$ by random sampling from $AllOps$. Thereafter, the constraints can be specified.
To model constraints as a CSP as $P_x(V, D, C)$, we define a variable set $V_{x}$ which contains all the numerical items from $Attr_x$ and $Input_x$, i.e., $V_{x} = Attr_x \cup \bigcup_{t \in Input_x} Str_{t}$, and also, specify the semantic specification of the operation $Spec_{OP_x}(Attr_x, Input_x)$ as a set of constraints $C$ on the variables. A specific example is shown in Equation~\ref{c1}-\ref{c3}.

To solve the above CSP, based on constraint propagation~\cite{apt2003principles}, our approach works as follows: iteratively extending the instantiation by picking an unassigned variable and assigning the value to the variable from its domains; after each turn's instantiation on a variable, with the help of the solver, our approach will conduct constraint propagation among unassigned variables throughout the constraints to reduce the domain of unassigned variables.

During the above process, there are three main issues as follows.

First, the existence of quantifiers and unknown functions in the constraints introduces the difficulty of conducting constraint propagation. For example, in Equation~\ref{c2}: $C_2: \forall i \in [1, 2, ..., d_x] \wedge (i \neq axis), x_i = y_i$, without instantiating the value of $axis$, we do not know whether $C_2$ implies $x_1=y_1$.

Second, how to instantiates $\psi_G$ (the structure of the graph) during the solving process. A straightforward way (in a `generate-then-match' style) works as follows: first instantiate all symbol variables in $Input_{x}$,
and then matches each tensor in $Input_{x}$ with instantiated tensors (i.e., outcoming edges of instantiated nodes) by comparing their structures.
However, because the  domain of the tensors under instantiation is large, the probability of exactly matching instantiated tensors is fairly small. For example, in the example of Section~\ref{sec:s2}, to instantiate tensor $x$ and $y$ (i.e., incoming edges of \CodeIn{Concat}), the number of possible solutions is exploded, which leads to fairly low probability of the equivalence between $Str_x$/$Str_y$ and structures of instantiated tensors (such as $Str_p$).
Thus, this straightforward way will lead to the result that generated computation graph tends to be simple and scattered.

Third, constraint propagation might produce an empty domain, causing backtracking of the solver (i.e., inconsistence of the instantiation), which affects the effectiveness of constraint solving. For example, for the constraint $(x=y) \wedge (x=z) \wedge (y \neq z)$, if the instantiation is $\left \langle x=1 \right \rangle$, after constraint propagation, $y$ holds an empty domain.


To address those issues, we tailor a well order that successfully
(1) simplifies the constraints to be quantifier-free and unknown-function-free, and
(2) enables to controllably select choices for instantiating unassigned tensors;
with the guarantee that our propagation-based instantiation is backtrack-free (keeps consistency during the whole process).

In the following parts, we will first describe the order, and then explain our reasons for that, finally, illustrate the property of \ttwo{} in our approach.

\subsubsection{The order of the instantiation}




For an operation $x$, our order contains several groups, arranged as follows: $G_1, G_2, G_3, ..., G_{|Indegree_x|+2}$. First group consists of one variable: $G_1=\{Indegree_x\}$. Second group is the variables related to the first tensor in the input (first incoming edge of this operation), i.e., $G_2 = Str_{in_1} (in_1 \in Input_x$).
Third group is the attributes of operations (except for $Indegree_x$), i.e., $G_3 = Attr_x \setminus \{ Indegree_x \}$.
For the rest, each group is variables related to the next tensor in the input (next incoming edge of this operation), i.e., $G_{i+2} = Str_{in_i} (in_i \in Input_x)$.
In the groups related to each tensor (assume the tensor is $t$), the variable that denotes the dimension size of the tensor is ahead of the variables that denote the length of each dimension of the tensor, i.e., $d_t$ (denotes $dim_t$) is ahead of $t_i$ (denotes $t[i]$).







\subsubsection{Reason I: Quantifier and unknown function elimination}\label{sec:s3.3.1}
Specifically, there are two forms which cause the existing constraint solving or sampling approaches~\cite{DBLP:conf/tacas/MouraB08, DBLP:conf/iccad/DutraBS18} hard to handle. First is the quantifier in the constraints, which hold the forms as $\forall i \in f(x), C(i) $, where $f$ is a function returns a set, $x$ is a variable, $C(i)$ is a constraint whose form is dependent on the value of $i$. Second is the unknown function. Constraints may contain terms such as $t_{f(x)}$, where $f(x)$ is a function that returns an integer number and $t_{f(x)}$ is the variable that denotes the $t[f(x)]$ (f(x)-st dimension's length of tensor $t$).

Our instantiation could eliminate the quantifiers and unknown functions in constraints with the following reason.
As quantifiers in form of $\forall i \in f(x), C(i)$, for all of variable $v$ whose existence in $C(i)$ depends on the domain of $f(x)$, we call that $v$ depend on $x$.
As constraints with unknown function such as $t_{f(x)}$, we call $t_i$ depends on $x$. 
For constraints on deep learning models, the above dependencies among constraints are loop-free. Our instantiation order is actually a topological order satisfying those dependencies, i.e., ensuring that for any dependencies that $x$ depends on $y$, $y$ is ahead of $x$ in the order. Thus, followed by the instantiation order, with satisfying the precondition of eliminating quantifiers and unknown functions by instantiation, we could simplify the constraints to a decidable propositional logic for constraint propagation.

\subsubsection{Reason II: On-demand instantiation}

The reason why we put the variables related to the same tensor in a group, i.e., instantiate the tensor(s) one by one, is due to the consideration of instantiating $\psi_G$.


We design an on-demand policy for instantiating the tensor, with consideration of the instantiation for $\psi_G$ in the meanwhile. For each tensor $t$ in the input of $x$ ($t \in Input_{x}$), 
our on-demand policy instantiates $Str_t$ with two choices: (1) reusing the structure $Str_s$ of an existing tensor $s$ as long as they are consistent with the instantiation, in other words, we set $t$'s structure the same as an instantiated tensor $s$, which is from the output of an instantiated node $ex$, i.e., instantiating $t = s$ with $\psi_{G}(t) = (ex, x)$ (if there are more than one satisfied tensor, we will randomly pick one of them; if no such tensor, we choose the second way); (2) creating a new tensor placeholder as a node $n$, and set its output as $t$, in other words, instantiating a tensor $t$ with $t \in Output_n$ as well as $\psi_G(t) = (n, x)$.

We select the choice of instantiating tensors according to a Bernoulli distribution, as a common way to produce random boolean decisions. Any other distributions are also allowed. The distribution is controlled by a parameter $picking\ rate$. The higher $picking\ rate$ is, the higher chance that our approach would select the first choice (i.e., reusing an existing tensor). To favor generating more complicated computation graphs, we set $picking\ rate$ relatively high in practice.
We will further explain the effect of this parameter in Section~\ref{sec:evaluation}.

\subsubsection{Property of operation-level global consistency}
We illustrate that our propagation-based instantiation will not lead to inconsistence due to the property we called operation-level global consistency.
For the constraints related to each single operation such as $x$, after determining the type ($OP_x$) and the number of tensors in the input ($Indegree_x$), we could take attributes as a variable and each tensor in the input of the operation as a variable. With good properties of specification on deep learning operations, this CSP is globally consistent~\cite{dechter1992local}. Thus, any order for instantiation, including our delicately-designed instantiation order in our approach, will not lead to inconsistence.

\subsection{Properties of \toolname{}}
Overall, \toolname{} in our approach is backtrack-free, sound and complete.
Based on the property of \firstp{} and opertional-level global consistency, we are able to instantiate without backtracking. And also, soundness is guaranteed because the instantiation is always consistent, leading to the satisfaction of final solutions.
Completeness is due to that we do not lose probability of generation of any instantiation during the whole process, in other words, any instantiation that satisfies the constraints has the possibility to be generated.

\section{Evaluation}\label{sec:evaluation}
\pre{To evaluate the effectiveness of our approach, we compare our approach with four baselines, including two state-of-the-art approaches Muffin~\cite{DBLP:conf/icse/GuLZ022} and TVMFuzz~\cite{DBLP:conf/sigsoft/ShenM0TCC21}.}
~\rev{To evaluate the effectiveness of our approach, we compare our approach with five baselines, including three state-of-the-art approaches Muffin~\cite{DBLP:conf/icse/GuLZ022},  TVMFuzz~\cite{DBLP:conf/sigsoft/ShenM0TCC21}, and NNSmith~\cite{NNSmith}.}
Also, we evaluate them on three popular real-world deep learning compilers to investigate their bug detection capability.
We construct the computation graph based on the ONNX~\cite{onnx} standard. Our implementation is on Python 3, supporting generation of \opnumber{} operations in total ~\cite{israproject}.
We address the following three research questions with an evaluation conducted on Ubuntu 20.04 of a laptop computer with Intel® Core™ i5-1135G7 CPU @ 2.40GHz and memory of 8GB. More details are included on our project website~\cite{israproject}.

RQ1: Is \toolname{} effective for generating test inputs for testing deep learning compilers?

RQ2: How effective and practical are the generated tests in revealing bugs in popular real-world deep learning compilers?

RQ3: Does our approach outperform state-of-art approaches in terms of testing deep learning compilers in terms of coverage and bug detection capability?

\subsection{Compared Work}

To assess the effectiveness of \toolname{}, we first design and implement two baselines as the representative of another two types of test generation techniques, named \boneshort{} and \btwoshort{}. In addition, we also compare \toolname{} with \pre{two} \rev{three} state-of-the-art approaches that can be applied for testing deep learning compilers.
More specifically, we include the following representative techniques in our evaluation:

\subsubsection{\boneshort{}}
Declarative-style generation constructs deep learning models only based on the syntax grammar, in short as \boneshort{}. When determining the shape of tensors, it just randomly generates from all choices.
After the construction of the input, i.e., a whole computation graph, this approach directly feeds input into the compiler under test, and relies on the compiler's running to check whether the model is satisfied with its semantic specifications.

\subsubsection{\btwo{}}
Inspired by feedback-directed random test generation~\cite{pacheco2007feedback}, this approach conducts random generation for operation construction, i.e., randomly constructs a new operation to append it into the model, and checks whether the model satisfies the semantic specifications or not. This generation way can avoid generating invalid models at early stages, leading to the improvement of overall effectiveness, while the generation for a single operation to satisfy its semantic specifications is still ineffective.

\subsubsection{Muffin}
Muffin~\cite{DBLP:conf/icse/GuLZ022} is a state-of-the-art generation-based testing approach, proposed originally to test deep learning libraries such as Keras~\cite{keras}, generating models based on two model structure templates (chain structure with skips and cell-based structure) with deep learning library APIs. To satisfy with tensor structural constraints, Muffin hardcodes additional \CodeIn{Reshaping} layers to reshape the input/output tensors for the connection between layers.
\rev{Though Muffin is not designed for testing deep learning compilers, we can retrofit it to barely achieve the goal by converting the models constructed by high-level APIs into computation graphs with existing tools~\cite{tf2onnx}.}

\subsubsection{TVMFuzz}
\pre{To our knowledge, TVMFuzz~\cite{DBLP:conf/sigsoft/ShenM0TCC21} is the only existing work specifically targeting testing deep learning compiler.}
\rev{TVMFuzz~\cite{DBLP:conf/sigsoft/ShenM0TCC21} is a fuzzer which specifically targets testing TVM~\cite{DBLP:conf/osdi/ChenMJZYSCWHCGK18}.}
It randomly generates and mutates Tensor-level IR(TIR) expressions (the intermediary language of TVM) for fuzzing test, mainly towards the type-related bug detection. 

\rev{
\subsubsection{NNSmith}
NNSmith~\cite{NNSmith} is a generation-based approach for testing deep learning compilers, as a parallel research work with ours. 
NNSmith constructs the computation graphs based on the specifications of deep learning models: it tries possible choices for types, attributes, and input/output of operations by iteratively adding constraints from specifications and leveraging existing constraint solver Z3~\cite{DBLP:conf/tacas/MouraB08} to check the satisfaction; if constraints are not satisfied, it will backtrack and try different choices to pick. 
}


\subsection{Study Subjects and Settings}~\label{s4.2}
We choose TVM \pre{and}\rev{,} Glow (two popular open-source compilers), and SophGo (a state-of-practice commercial compiler) as our study subjects.
For TVM and Glow, we download their official released versions from GitHub: TVM as v0.7 (commit id 728b829), Glow\footnote{Glow does not have release version on GitHub. Thus we directly download its latest version.} (commit id a2036bd). For SophGo, we attain its latest released version from the company that develops it.
%



%
For test oracles, we use two types of oracles: (1) runtime failure (including error/crash behaviors) of compilers, i.e., when a computation graph causes the exception of compilers (excluding invalid inputs that violate the specifications);  (2) differential testing by feeding the same random inputs and comparing the outputs of compiled models from different compilers.
In differential testing, 
we set the relative error as 10\% (we set this value relatively large in order to avoid many false positives) for automatically comparing results from different compilers under the same input. 




For \toolname{}, we set the upper bound on the tensor dimension as 5, and the upper bound on the length of each dimension as 5 (which is aligned with default settings of Muffin). For $picking\ rate$, we set it with $0.97$ based on our sensitivity experiments, as shown in Table~\ref{tab:pickrate}. Except for above parameters which keep the same among all of experiments, we set the lower bound and upper bound of operation number in the graphs (named $lb$ and $ub$) according to the experiments, the $numop_G$ is uniform sampling between $lb$ and $ub$. 

For Muffin, we obey its default settings. For alignment with comparsions on coverage metrics, we convert the Keras models generated by Muffin to ONNX format with TF2ONNX~\cite{tf2onnx}. Because Keras API is more higher level than ONNX, the converting leads to the growth of model size. For fairness, in the experiment on coverage measurement, for every model Muffin generated, we adopt \toolname{} to  generate a model with the number of operations same as Muffin, named \toolname{}*. Also, we ensure that both approaches have a same operation set (of size 36, which is the number of overlapped operations).



For TVMFuzz, we obey its default parameters. Due to the difference on the type of generated inputs (\toolname{} generates computation graphs, while TVMFuzz generates programs in TIR, an IR for the TVM compiler), the metrics that we design for the work in this paper are not applicable for TVMFuzz. So we are unable to measure our coverage metrics on TVMFuzz. TVMFuzz is compiler-dependent, so we are unable to test TVMFuzz on compilers except for TVM.

\rev{For NNSmith, we obey its default parameters. For fairness, in the experiment on coverage measurement, we align the settings of \toolname{} with those from NNSmith, named \toolname{}**. For every model NNSmith generated, \toolname{}** generates a model with the same number of operations. Also, we ensure that both approaches have a same operation set (of size 40, which is the number of overlapped operations).}

For the experiment on coverage measurement, we run each approach to generate a fixed number of models, as 10000. For the parameters of operation number in the graphs, we set $lb$ as 1 and $ub$ as 200 for \toolname{} as well as \boneshort{} and \btwoshort{} in the experiment.
To eliminate randomness, We run our approach and baselines separately for five times and calculate the average number.
According to our results of the experiments, the coverage metrics of the three approaches are all saturated or converged, so it is a reasonable setting for our evaluations. 

For investigating bug detection capability, aligned with the setting in ~\cite{DBLP:conf/icse/GuLZ022}, for each method, we generate 300 computation graphs for testing compilers. Due to the difference of supported operations for each compiler, we run the generators with filtering the generated graphs that contain unsupported operations until the graph number reaches to 300. To reduce manual work on deduplication, we set the generated model relatively small with $lb$ as 1 and $ub$ as 10.


To save manual effort for checking duplicated bugs, we automatically check and filter the bugs with the same error messages and stack traces as produced before. For the rest of the bugs, two authors manually check their root causes through error messages, stack traces, and generated inputs to further eliminate duplicated bugs and false positives.


\subsection{Metrics}
In order to evaluate the effectiveness and practicability of different approaches, we investigate on their bug-detection capability, by counting the number of distinct bugs they detect within a fixed number of test inputs when used to test real-world deep learning compilers.

Furthermore, to better measure various aspects of generated inputs, 
\rev{we inherit the coverage criteria from previous research work~\cite{DBLP:conf/icse/LuoCRWFC21} and expand upon them to measure the diversity of computation graphs, including types, tensor shapes, and attributes of operations as well as connections between operations.
The design of these coverage criteria is motivated by the
fact that (1) existing traditional code coverage criteria are ineffective in deep learning testing~\cite{DBLP:conf/icse/LuoCRWFC21}, and neuron coverage criteria~\cite{DBLP:conf/sosp/PeiCYJ17} are also invalid in our evaluation because compiler testing involves different input models, and (2) type and shape problems are major root
causes of deep learning compiler bugs as demonstrated in a recent study~\cite{DBLP:conf/sigsoft/ShenM0TCC21}, also, (3) a lot of high-level optimizations in deep learning compilers, which are error-prone due to continuous and frequent development and modifications~\cite{DBLP:conf/sigsoft/ShenM0TCC21, du2021empirical}, could only be triggered by different types of specific connections between operations in the computation graphs~\cite{HirGen}.
}


\rev{Specifically,} we design 11 types of metrics for measuring coverage among input space.
The metrics are of two major categories: graph-level metrics and operation-level metrics. 
For operation-level metrics, we mainly follow the work by ~\cite{DBLP:conf/icse/LuoCRWFC21}.
For graph-level metrics, we design them with analogy of concepts in structural code coverage and combinatorial testing. Besides the preceding metrics, we also count the frequency of occurrences for operations and calculate the distribution of operations. 


\subsubsection{Graph-level metrics}
\rev{Graph-level metrics are designed for measuring various aspects of a single generated model.}
Let a \textbf{test set} be the set of models generated by an  approach. Given a test set $I$, which contains $n_I$ graphs, the graph-level metrics are defined as follows. 



\textbf{Number of Operations (NOO)} of a graph $g$ is defined as the total number of operations in $g$. 
\textbf{Number of Operation Types (NOT)} of a graph $g$ is defined as the number of different types of operations in $g$. 
\textbf{Number of Operation Pairs (NOP)} of a graph $g$ is defined as the number of different edges in $g$ that connect two operations together. 
\textbf{Number of Operation Triples (NTR)} of a graph $g$ is defined as the number of different pairs of adjacent edges that connect three operations together in $g$.   
\textbf{Number of Shapes and Attributes (NSA)} of a graph $g$ is defined as the total number of different shapes of input tensors and attributes (except for its input dgrees) for each operation in graph $g$. 
These graph-level metrics for test set $I$ are defined as the average of each of them among all the graphs in $I$: $GLM(I) = \frac{\Sigma_g{GLM_g(g)}}{n_I}$, where $GLM \in \{NOO, NOT, NOP, NTR, NSA\}$.

\subsubsection{Operation-level metrics}
\rev{Operation-level metrics are designed for measuring various aspects of operations among the whole test set.}
An \textbf{operation corpus} is a set of operations with their attributes including the operation name and the possible number of different input degrees. 
Given an operation corpus $C$ containing $n_C$ different operations and a test set $I$, 
we first calculate the metric on each type of operator $o$ in $I$, denoted as $XXC_{op}(o)$, then we have the operation-level metric of $I$ as the average of the operation-level metric on different operators, i.e., $XXC(I)=\frac{\Sigma_o{XXC_{op}(o)}}{n_C}$, where $XXC\in\{OTC,IDC,ODC,SEC,DEC,SAC\}$. We simply explain the meanings of these six metrics as below, and detailed definitions of these operation-level metrics are shown in our project website~\cite{israproject}.
  
\textbf{Operation Type Coverage (OTC)}, \textbf{Input Degree Coverage (IDC)}, \textbf{Output Degree Coverage (ODC)} show the diversity of operations types, and the diversity of the input and output degrees of them in the test set $I$ respectively. 
\textbf{Single Edge Coverage (SEC)} shows the diversity of edges between the operations in $I$.
\textbf{Double Edge Coverage (DEC)} shows the diversity of pairs of edges that are adjacent, which is actually the coverage of different triples of operations that are connected in a graph in the test set.
\textbf{Shapes and Attributes Coverage (SAC)} indicates the diversity of attributes of the operations (except for input degrees) and their input tensor shapes in the test set.

\begin{table*}[]

    \caption{Results on Graph-level and Operation-level Metrics}
    \label{tab:1_200}
    
    \centering

\begin{tabular}{c|rrr|rr|rr}
\textbf{}       & \multicolumn{1}{c}{\textbf{Isra}} & \multicolumn{1}{c}{\textbf{DeclGen}} & \multicolumn{1}{c|}{\textbf{\begin{tabular}[c]{@{}c@{}}Randoop-\\ Gen\end{tabular}}} & \multicolumn{1}{c}{\textbf{Isra*}} & \multicolumn{1}{c|}{\textbf{Muffin}} & \multicolumn{1}{c}{\rev{\textbf{Isra**}}} & \multicolumn{1}{c}{\rev{\textbf{NNSmith}}} \\ \hline
\textbf{time/s} & \textbf{320.1713}                 & 9011.9172                            & 4721.8233                                                                            & \textbf{20.1031}                   & 25847.6804                           & \rev{\textbf{111.8615}}                   & \rev{880.7737}                             \\
\textbf{}       &                                   &                                      &                                                                                      &                                    &                                      &                                     &                                      \\
\textbf{OTC}    & \textbf{100\%}                    & 97.536\%                             & 98.46\%                                                                              & \textbf{100\%}                     & \textbf{100\%}                       & \rev{\textbf{100\%}}                      & \rev{\textbf{100\%}}                       \\
\textbf{IDC}    & \textbf{92.95\%}                  & 90.178\%                             & 89.966\%                                                                             & \textbf{91.85\%}                   & 88.52\%                              & \rev{\textbf{89.54\%}}                    & \rev{88.71\%}                              \\
\textbf{ODC}    & \textbf{11.848}                   & 4.928                                & 10.616                                                                               & \textbf{8.75}                      & 4.22                                 & \rev{\textbf{6.925}}                      & \rev{6.225}                               \\
\textbf{SEC}    & \textbf{98.27\%}                  & 67.804\%                             & 88.08\%                                                                              & \textbf{98.15\%}                   & 35.49\%                              & \rev{\textbf{99.38\%}}                    & \rev{78.50\%}                              \\
\textbf{DEC}    & \textbf{90.208\%}                 & 2.126\%                              & 45.324\%                                                                             & \textbf{57.7\%}                    & 4.95\%                               & \rev{\textbf{84.30\%}}                    & \rev{28.70\%}                              \\
\textbf{SPC}    & \textbf{3001.938}                 & 227.018                              & 1509.356                                                                             & \textbf{1192.22}                   & 556.44                               & \rev{\textbf{2822.9}}                     & \rev{1641.725}                             \\
\textbf{}       &                                   &                                      &                                                                                      &                                    &                                      & \multicolumn{1}{l}{}                & \multicolumn{1}{l}{}                 \\
\textbf{NOO}    & \textbf{100.8766}                 & 2.8783                               & 100.1231                                                                             & \textbf{10.4236}                   & \textbf{10.4236}                     & \rev{\textbf{16.3999}}                    & \rev{\textbf{16.3999}}                     \\
\textbf{NOT}    & \textbf{45.237}                   & 2.76                                 & 33.0021                                                                              & \textbf{8.6589}                    & 5.3289                               & \rev{\textbf{13.1333}}                    & \rev{10.5113}                              \\
\textbf{NOP}    & \textbf{103.7621}                 & 1.4856                               & 98.6460                                                                              & \textbf{7.8243}                    & 6.399                                & \rev{\textbf{14.5851}}                    & \rev{10.4880}                              \\
\textbf{NTR}    & 102.9130                          & 0.6042                               & \textbf{105.5774}                                                                    & 4.6481                             & \textbf{6.0766}                      & \rev{\textbf{13.065}}                     & \rev{7.6740}                               \\
\textbf{NSA}    & \textbf{26.6252}                  & 1.5533                               & 10.0211                                                                              & 5.9253                             & \textbf{11.3604}                     & \rev{\textbf{10.8192}}                    & \rev{10.1459}                             
\end{tabular}


\end{table*}

\subsection{RQ1: Evaluation Results of Generated Inputs}


\subsubsection{Operation-level metrics}
The result of experiment on coverage measurement is shown in Table~\ref{tab:1_200}. Firstly, with alignment on the number of generated inputs,
we can find that \toolname{} outperforms the baselines greatly with respect to the amount of time, showing the efficiency of our approach.

For operation-level metrics, we find that \toolname{} is able to cover all kinds of operations that we have chosen and all kinds of input degrees of each type of operation. Compared with the two baselines,  \toolname{} has higher coverage on all of operation-level metrics, especially for $SEC$, $DEC$, and $SAC$.

\subsubsection{Graph-level metrics}
We find that the $NOO$ (Number of Operations) of our approach and \btwoshort{} are closer to the average of the lower bound and upper bound of the operation numbers that we set (consistent with our uniform sampling for the number of operations), while
\boneshort{}'s $NOO$ remains at a significantly lower level. The reason is that \boneshort{} holds less probability to satisfy the semantic specifications, which leads to generating rather simple models and bad performance on the graph-level metrics. 
\btwoshort{}'s $NOP$ and $NTR$ are comparable with our approach, however, the operation types ($NOT$) of \toolname{} are more diverse, making it outperform \btwoshort{} at coverage of operation pairs ($SEC$) and triples ($DEC$). 
The results of graph-level metrics indicate that \toolname{} are capable of generating diverse, large and complex models.

\subsubsection{\pre{Diversity} \rev{Distribution of Operation Frequency}}
\begin{figure}
    \includegraphics[width=\linewidth]{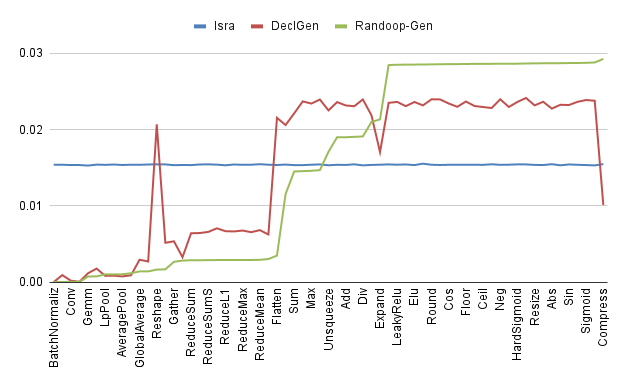}
    \caption{Distribution of Operation Frequency of \toolname{}  \pre{and the baselines} \rev{, \boneshort{}, and \btwoshort{}} (captioning only parts of operations in x-axis for a clear presentation, numbers at y-axis are normalized with the total number). \pre{Muffin is not included due to the difference on the supported operation set.}}
    \label{fig:chart}
\end{figure}

\begin{figure*}
\vspace{-4mm}
\begin{minipage}{\textwidth}
    \begin{minipage}[]{0.49\textwidth}
    \includegraphics[width=\linewidth]{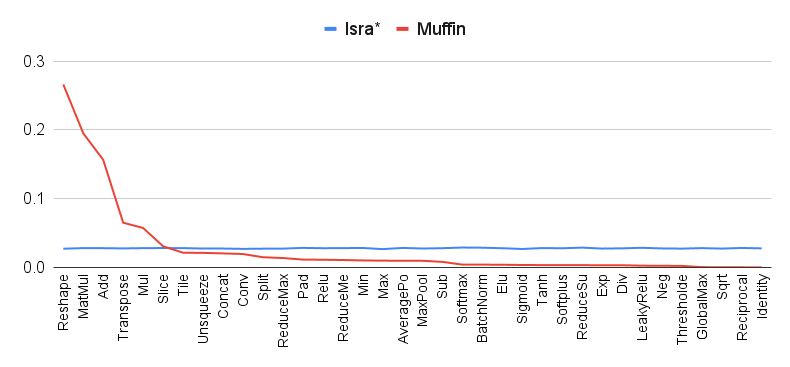}
    \vspace{-7mm}
    \caption{\rev{Distribution of Operation Frequency of Isra* and Muffin.}}
    \label{fig:withMuffin}
    \end{minipage}
    \hfill
    \begin{minipage}[]{0.49\textwidth}
    \includegraphics[width=\linewidth]{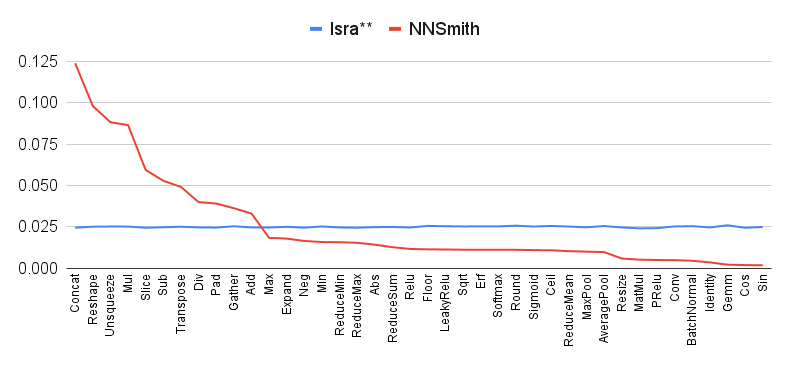}
    \vspace{-7mm}
    \caption{\rev{Distribution of Operation Frequency of Isra** and NNSmith.}}
    \label{fig:withNNSmith}
    \end{minipage}

\end{minipage}


\vspace{-5mm}
\end{figure*}

As shown in Figure~\ref{fig:chart}, \rev{Figure~\ref{fig:withMuffin}, and Figure~\ref{fig:withNNSmith},}
\toolname{} is able to generate different operations in a relatively uniform distribution, which is consistent with our uniform sampling for picking the type of operation. As a contrast, \pre{both of \boneshort{} and \btwoshort{}}\rev{all of baselines} fail to generate a sufficient number of operations with relatively complicated semantic specifications such as \CodeIn{Conv} and \CodeIn{Gemm}. It shows that \pre{\boneshort{} and \btwoshort{}} \rev{all of baselines} have a limitation that the diversity of models generated by them is weak. This is because the constraints of some operations are relatively
complicated and less possible to satisfy. Those complex operations are less possible to be chosen in the valid output by the filter of the \rev{iterative} checking process. \rev{Besides, among models generated by Muffin,
\CodeIn{reshape} operation appears most frequently because Muffin forces the insertions of \CodeIn{reshape} operation before many operations to ensure semantic specifications.}

\begin{table}
      \centering 

		\caption{Results of Sensitivity Experiment of $picking\ rate$}
        \label{tab:pickrate}
        \begin{tabular}{crrrr}
                \multicolumn{1}{c}{\textbf {\begin{tabular}[c]{@{}c@{}}Picking-\\ Rate\end{tabular}}}
                & \multicolumn{1}{c}{\textbf{time/s}} & \multicolumn{1}{c}{\textbf{NOP}} & \multicolumn{1}{c}{\textbf{NTR}} & \multicolumn{1}{c}{\textbf{NSA}} \\
                \textbf{0.5}          & \textbf{215.29}                 & 51.07                         & 23.73                          & \textbf{62.86}                \\
                \textbf{0.8}          & 260.47                          & 82.91                          & 63.99                         & 42.05                         \\
                \textbf{0.9}          & 269.98                          & 94.34                         & 83.30                         & 33.84                         \\
                \textbf{0.95}         & 270.01                          & 100.17                         & 94.19                         & 29.54                         \\
                \textbf{0.97}         & \textbf{267.24}                 & \textbf{102.56}               & \textbf{98.77}                & \textbf{27.77}                \\
                \textbf{0.99}         & 266.36                          & \textbf{104.94}               & \textbf{103.57}               & 25.93                        
        \end{tabular}

	

\end{table} 


\subsubsection{$picking\ rate$} To evaluate the effect of $picking\ rate$ parameter, we also compare the results of \toolname{} with the setting of different $picking\ rates$. The settings are the same as the experiment on coverage measurement, except that we set the operation number in the graphs as a fixed number 100.
The result is shown in Table~\ref{tab:pickrate}.
If the $picking\ rate$ is relatively high, the operations are more likely to be matched with existing tensors, leading to $NOP$, $NTR$ going high. In the meanwhile, the shape of newly created tensors is more likely to be equal to the shape of previous tensors, leading to the result that $NSA$ goes down as the $picking\ rate$ increases. We finally choose its value as 0.97, as a trade-off.





\begin{figure*}
\vspace{3mm}
    \begin{minipage}[]{0.53\textwidth}
      \captionof{table}{\# unique bugs detected on different compilers.}
      		\centering
        \tabcolsep=0.4mm
        \label{tab:bug_number}
        \renewcommand\arraystretch{2}
        \begin{tabular}{ccccccc}
            & \toolname{} & \boneshort{} & \multicolumn{1}{c}{\begin{tabular}[c]{@{}c@{}}Randoop-\\ Gen\end{tabular}} & Muffin & TVMFuzz & \rev{NNSmith\footnotemark}  
            \\
            TVM        & \textbf{18}      & 8                            & 14                      & 12     & 5  & \rev{\textbf{21}}     \\
            Glow       & \textbf{4}       & 2                            & 2                       & 2      & -  & \rev{\textbf{4}}     \\
            SophGo     & \textbf{11}      & 3                            & 3                       & 9      & -  & \rev{7}     \\
            Total      & \textbf{33}      & 13                           & 19                      & 23     & 5  & \rev{\textbf{32}}    
        \end{tabular}

    \end{minipage}
  \hfill
  \begin{minipage}[]{0.4\textwidth}
    \centering
    \includegraphics[width=\linewidth]{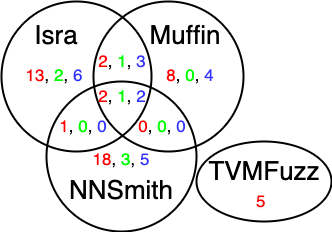}
    \captionof{figure}{\rev{The overlaps of detected bugs among different approaches on testing three compilers (red for TVM, green for Glow, blue for SophGo).}}
    \label{fig:overlap}
  \end{minipage}
\end{figure*}




\subsection{RQ2: Evalution Results on Bug-Detection Capability}

We investigate the effectiveness of \toolname{} and baselines in terms of distinct bugs detected in the same version of compiler subjects. 
After analysis of deduplication, \toolname{} detects \bugnumber{} unique bugs in total, \rev{as shown in Table~\ref{tab:bug_number}, } \pre{outperforming} \rev{performing better or as well than} baselines on all of three benchmarks.
The capability on bug detection shows the high effectiveness of \toolname{}.
\pre{The details of all of the detected bugs can be found on our project website~\cite{israproject}.}
\rev{To further evaluate bug-detection capability of \toolname{}, we conduct a study on the bugs \toolname{} detected.}



\addtocounter{footnote}{0}
\footnotetext{The number of NNSmith's detected unique bugs on TVM is different from the results in the NNSmith's paper~\cite{NNSmith} due to the different settings on the compiler versions and testing amount.}

\rev{
\subsubsection{Bug Study}
}

\pre{In addition, we categorize the bugs found by \toolname{} as two types} \rev{Among all of unique bugs detected by ~\toolname{}, we categorize them into two types based on the test oracles detecting them}: (1) \emph{error bug} (29 of \bugnumber{}) : the given input triggers an \rev{error,} crash or unexpected exception during compilation; (2) \emph{inconsistency bug} (4 of \bugnumber{}): the compiler executes the given input and terminates normally but the compiled model produces different outputs according to differential testing.

After we report found bugs to the corresponding community, compiler developers give responses for most of them, all with positive feedback.
Among all of bugs found by \toolname{}, there are \unknownbug{} previously unknown bugs.
Until now, a majority of our detected bugs (27 of \bugnumber{}) have already been fixed by developers, with 16 bugs directly owing to our bug reporting/pull requests, benefiting all the compiler users and their applications. 
\rev{One of TVM core developers replies with the following message for bugs reported by us\footnote{\url{https://github.com/apache/tvm/pull/7208\#issuecomment-754406762}}: ``These two errors that you generated were excellent real bugs with the importer and were very easy to understand and replicate with your post. If they're being auto-generated they look excellent!''}
The feedback from real-world developers is also strong evidence showing that \toolname{} is practical for testing real-world deep learning compilers and able to detect critical bugs in practice. 



\rev
{
We list 24 previously unknown bugs (as well as bug reports) detected by \toolname{} in Table~\ref{unknownbugs}, including the stage of root causes, the time it takes to have the bug confirmed and fixed, and the GitHub issue ID~\footnote{For reference, the URLs of TVM's and Glow's bug reports are https://github.com/apache/tvm/issues/\#IssueID, and https://github.com/pytorch/glow/issues/\#IssueID. The URLs for SophGo's bug reports are internal and not publicly accessible.} for reference. The details of confirmed/fixed bugs are as follows.

\textbf{TVM-1} is a bug in the compiler backend, caused by a pass named \CodeIn{DynamicToStatic} which should not be defined at optimization level 3. It will lead the internal error when the deep learning model contains operations such as \CodeIn{MatMul}, \CodeIn{Dropout}. After our reporting, developers reordered passes in the backend and lowered \CodeIn{DynamicToStatic} optimization level to fix it~\footnote{\url{https://github.com/apache/tvm/pull/7213}}.

\textbf{TVM-2} is a bug in the compiler frontend. The TVM developer has explained the cause of the bug and fixed it\footnote{\url{https://github.com/apache/tvm/issues/7202\#issuecomment-754372403}}: ``It's due to a bad assumption made in \CodeIn{PRelu} conversion about the input layout......Our current PRelu converter assumes that incoming data is in NCHW format and that the slope will have C total elements. Neither of these are actual requirements for ONNX \CodeIn{PReLu}.''

\textbf{TVM-3} and \textbf{TVM-4} are bugs in the compiler frontend. Some of parameters in \CodeIn{LpPool} and \CodeIn{LeakyRelu} operation in ONNX standard allow default value but it was not supported in TVM.

\begin{table}
\centering

\rev{

\setlength{\tabcolsep}{2pt}

\caption{\rev{Previously unknown bugs detected by \toolname{}.}}
\label{unknownbugs}
\begin{tabular}{cccccc}
ID        & Type         & Location & Confirmed        & Fixed            & Issue ID \\
TVM-1     & error        & backend  & \textless 1 day  & 10 days          & 7200     \\
TVM-2     & error        & frontend & \textless 1 day  & \textless 1 day  & 7202     \\
TVM-3     & error        & frontend & \textless 1 day  & 6 months         & 7241     \\
TVM-4     & error        & frontend & \textless 1 day  & \textless 1 day  & 7244     \\
TVM-5     & error        & backend  & 7 months         & 7 months         & 7262     \\
TVM-6     & error        & frontend & 42 days          & 42 days          & 7263     \\
TVM-7     & error        & frontend & \textless 2 days & 10 months        & 8889     \\
TVM-8     & error        & backend  & \textless 2 days & 10 months        & 8890     \\
TVM-9     & inconsistency & backend  & N/A              & 10 months        & 7270     \\
TVM-10    & inconsistency & unknown  & N/A              & N/A              & 12734    \\
TVM-11    & error        & unknown  & N/A              & N/A              & 12735    \\
Glow-1    & error        & backend  & 13 days          & N/A              & 5995     \\
Glow-2    & error        & unknown  & N/A              & N/A              & 5991     \\
SophGo-1  & error        & frontend & \textless 3 days & N/A              & *        \\
SophGo-2  & error        & backend  & \textless 3 days & \textless 7 days & *        \\
SophGo-3  & error        & backend  & \textless 3 days & \textless 7 days & *        \\
SophGo-4  & error        & frontend & \textless 3 days & \textless 7 days & *        \\
SophGo-5  & error        & backend  & \textless 3 days & \textless 7 days & *        \\
SophGo-6  & error        & backend  & \textless 3 days & \textless 7 days & *        \\
SophGo-7  & inconsistency & backend  & \textless 3 days & \textless 7 days & *        \\
SophGo-8  & error        & frontend & \textless 3 days & \textless 7 days & *        \\
SophGo-9  & error        & frontend & \textless 3 days & \textless 7 days & *        \\
SophGo-10 & error        & frontend & \textless 3 days & N/A              & *        \\
SophGo-11 & error        & backend  & \textless 3 days & \textless 7 days & *       
\end{tabular}

}



\vspace{-3mm}
\end{table}





\textbf{TVM-5} is a bug in the compiler backend, which catches an edge case of the compiler's type checker as explained by a TVM developer\footnote{\url{https://github.com/apache/tvm/issues/7262\#issuecomment-911968074}}.

\textbf{TVM-6} and \textbf{TVM-7} are bugs in the compiler frontend. The former is due to that one of input tensors in \CodeIn{Gemm} operation can be optional by the standard but the TVM doesn't allow that behavior. The latter is due to the fact that ``\CodeIn{Split} operation is not dynamic inputs'' as explained by the compiler developer.



\textbf{TVM-8} is a bug in the compiler backend due to inconsistent specifications between different versions on \CodeIn{Squeeze} operation, causing erroneous implementation in the compiler.



\textbf{TVM-9} is a bug in the compiler backend. The erroneous backend implementation causes the inconsistent outputs on a model containing \CodeIn{Flatten} and \CodeIn{ReduceL1} operation. Developers have not confirmed this bug, but it has been fixed in the next released version of the compiler.

\textbf{Glow-1} is a bug in the compiler backend due to that \CodeIn{ReduceSum} operation in the input model contains multi axis inputs and attributes.




\textbf{SophGo-1} is a bug in the compiler frontend. The compiler does not support that the weight tensor of \CodeIn{Conv} operation is an input tensor. Developers do not fix this bug due to the reason that they suppose users take this parameter as a constant because this parameter usually does not change after deployment.

\textbf{SophGo-2} is a bug in the compiler backend. If the inputs of \CodeIn{Mean} operation are the same tensor, then the calculation of mean operation will throw exception due to the naming error.

\textbf{SophGo-3} is a bug in the compiler backend. \CodeIn{ReduceProd} operation will register a buffer, but buffer size has not been assigned which leads to the incorrect parameters in calculation and further causes the final result wrong.

\textbf{SophGo-4} is a bug in the compiler frontend due to erroneous parsing for \CodeIn{Pooling} operation.

\textbf{SophGo-5} is a bug in the compiler backend due to incorrect implementation on \CodeIn{Split} operation.

\textbf{SophGo-6} is a bug in the compiler backend. The compiler assumes the second input of \CodeIn{PRelu} holds a specific shape which is not consistent with the specification.

\textbf{SophGo-7} is a bug in the compiler backend. The compiler backend incorrectly covers the input data for the calculation on \CodeIn{Cos} operation, causing the wrong results of model's output.

\textbf{SophGo-8}, \textbf{SophGo-9}, and \textbf{SophGo-10} are bugs in the compiler frontend due to erroneous parsing for 
\CodeIn{Unsqueeze} operation, \CodeIn{Split} operation and \CodeIn{Gemm} operation. ONNX standard of some operations has changed after version 13. The compiler only implements old version and is not compatible with latest standard. Developers do not fix \textbf{SophGo-10} due to the same reason as \textbf{SophGo-1}.

\textbf{SophGo-11} is a bug in the compiler backend due to incorrect implementation on \CodeIn{Resize} operation.
}

\rev{
\subsubsection{False positives}
Though our test generation approach is sound (i.e., the models generated by our approach ensures the validity), false positives may be introduced by floating point errors~\cite{7194603, DBLP:journals/pomacs/XiaoLYPW22}.
For differential testing, one of test oracles we used,
when it comes to checking output equivalence,
false alarms may be produced due to floating point errors/inaccuracies.
To reduce false alarms, we use a high error tolerance in comparison of models' outputs (the relative error is set to 10\% as mentioned in Section~\ref{s4.2}), which is similar to a parallel work~\cite{NNSmith}.
In our evaluation, \toolname{} did not
find false positives caused by the floating-point errors.
}

\subsection{RQ3: Comparison with State-of-the-art Approaches}

We compare Isra with \pre{two} \rev{three} state-of-the-art approaches \rev{(i.e., Muffin~\cite{DBLP:conf/icse/GuLZ022}, TVMFuzz~\cite{DBLP:conf/sigsoft/ShenM0TCC21} and NNSmith~\cite{NNSmith})} in terms of coverage and bug detection capability.

For the comparison on coverage, as result shown in Table~\ref{tab:1_200}, \pre{under the same settings, compared with Muffin, \toolname{}* outperforms on most of metrics except for only $NTR$ and $NSA$.} \rev{(1) under the same settings, \toolname{}* outperforms Muffin on most of coverage metrics except for only $NTR$ and $NSA$,
and (2) under the same settings, \toolname{}** consistently outperforms NNSmith on all of coverage metrics. The reasons why Muffin achieves higher $NTR$ and $NSA$ than \toolname{}* are as follows.}
For $NTR$, this is mainly because Muffin contains a cells mode, which will favor generating dense graph structure which contains many triples, but its $DEC$ is still significantly low compared with Isra*, due to the fewer types of operations in per graph ($NOT$). Muffin's higher coverage on $NSA$ is
because Muffin inserts \CodeIn{Reshape} layers between adjacent layers for patching tensor structural constraints in a hardcode way, leads to change tensor shape frequently. However, Muffin is still significantly lower than \toolname{} on $SAC$, which is the coverage of $NSA$ among the whole test set. 

For the comparison on bug detection capability, 
\rev{as shown in Table~\ref{tab:bug_number}}, \toolname{} detects more bugs than Muffin on three compilers, with 1.5x, 2x, 1.22x respectively, in total 33 versus 23. Also, \toolname{} significantly outperforms TVMFuzz on detecting bugs on TVM with 18 versus 5.
\rev{\toolname{} achieves comparable results compared to NNSmith (33 versus 32) on three compilers.
In addition, we also investigate the overlaps of detected bugs among \toolname{}, Muffin, TVMFuzz, and NNSmith as shown in Figure~\ref{fig:overlap}.
There are overlapping bugs among them as well as distinct non-overlapping bugs, indicating that these approaches are complementary.
There is no overlapping bug between TVMFuzz and other approaches, primarily because TVMFuzz fuzzes on low-level Relay IR instead of computational graphs.
}


\rev{
Overall, as shown in evaluation results, compared to state-of-the-art approaches under the same settings, \toolname{} outperforms those approaches on various coverage metrics, and also achieves comparable and complementary results on bug detection.
The results indicates that (1) \toolname{} is more effective and more efficient than existing approaches for test generation,
(2) \toolname{} is effective and complementary to existing approaches for detecting bugs in real-world compilers.


}

\section{Related Work}

\textit{Random Testing and Test Generation}. 
Random testing~\cite{orso2014software} simply constructs test inputs in a random manner. For example, Randoop~\cite{pacheco2007feedback} and EvoSuite~\cite{fraser2011evosuite} aim to generate JUnit tests for classes by incrementally combining sequences of method calls. Besides many aspects of advantages, random testing still faces problems including inefficiency, imprecision, and lack of available information to guide test generation.
To test deep learning compilers, our work conducts random testing by enhancing the effectiveness of test generation.
Another test generation way is bounded-exhaustive testing. For example, UDITA~\cite{gligoric2010test} uses bounded-exhaustive testing to enumerate the paths through the generator with various optimizations. For deep learning models, the space of computation graph and the shape of tensors in it can be super large, and the valid space is very sparse; thus, it is intolerable to enumerate all kinds of the inputs by searching.


\textit{Grammar-based Fuzzing}.
Fuzzing is a common approach for randomly generating inputs to test software.
It may generate inputs from scratch, or do mutation on a series of valid seed inputs. 
Without any knowledge of the input of the software under test, generating valid inputs randomly is ineffective, especially for the software such as compilers whose inputs are highly-structured.
To improve it, grammar-based fuzzing~\cite{godefroid2008grammar, DBLP:conf/uss/HollerHZ12} is proposed, which relies on a grammar specification to generate structured inputs, usually in context-free forms. Deep learning models with semantic specifications fail to be represented as a context-free grammar.
Recently Padhye et al.~\cite{padhye2019semantic} propose Zest, which is based on coverage-guided fuzzing, targeting at producing semantically valid inputs. 
However, Zest still requires developers to manually design a generator that can construct syntactically valid test programs. Different implementations for the generator could highly affect the effectiveness of test generation, especially for languages with complicated specifications such as deep learning models.

\rev{
\textit{Testing Deep Learning Toolkits}. 
Deep learning toolkits include deep learning libraries (frameworks) and deep learning compilers. The differences between them lie in their primary functions and interfaces provided to users, which result in divergent emphases in designing corresponding testing approaches.

Deep learning libraries, such as Keras and TensorFlow, are primarily used for simplifying the implementation of deep learning models, they provide high-level APIs and abstractions of pre-defined layers/models as well as optimizers, loss functions and other utilities that allow users to easily define and train deep learning models. To test deep learning libraries, LEMON~\cite{DBLP:conf/sigsoft/WangYCLZ20} and Muffin~\cite{DBLP:conf/icse/GuLZ022} focus on generating parameters and call sequences of high-level APIs.


Deep learning compilers, such as TVM, are designed to transform deep learning models into efficient low-level code for deployment and execution on different hardware devices,
and they focus on optimizing the computational graph of the models to improve execution efficiency.
To test deep learning compilers, one direction is to directly generate inputs of deep learning compilers, i.e., computation graphs, including research work GraphFuzzer~\cite{DBLP:conf/icse/LuoCRWFC21} and MT-DLComp~\cite{DBLP:journals/pomacs/XiaoLYPW22}; another direction is to fuzz low-level intermediate representation of the compiler (e.g., TVM's compiler-specific intermediate representation), including research work TVMFuzz~\cite{DBLP:conf/sigsoft/ShenM0TCC21} and Tzer~\cite{DBLP:journals/pacmpl/LiuWYDZ22}.
Our approach, \toolname{}, as well as its two parallel works NNSmith~\cite{NNSmith} and HirGen~\cite{HirGen} belong to the former, i.e. test generation of computation graphs.

}

\rev{
To generate test inputs for deep learning toolkits, the validity of test inputs is a critical challenge: invalid test inputs will largely diminish effectiveness and efficiency of testing.
To address it, different techniques are proposed by existing research work.
For example, Muffin~\cite{DBLP:conf/icse/GuLZ022}, GraphFuzzer~\cite{DBLP:conf/icse/LuoCRWFC21} and HirGen~\cite{HirGen} are all restricted to certain types of operations and connections for generating computation graphs, which will bias the generated graphs.
Specifically, Muffin~\cite{DBLP:conf/icse/GuLZ022} ensures the semantic specification by inserting the reshape layers between adjacent layers in origin models. Unfortunately, doing so biases the generated computation graphs to include many \CodeIn{Reshape} operations as shown in our evaluation.
GraphFuzzer~\cite{DBLP:conf/icse/LuoCRWFC21} and HirGen~\cite{HirGen} try to adjust mismatched tensor shapes through slicing and padding. They will also bias the generated computation graphs to include
many \CodeIn{Slice} and \CodeIn{Padding} operations.

NNSmith~\cite{NNSmith}, as a parallel work with ~\toolname{}, addresses the validity challenge by leveraging the existing constraint solver Z3~\cite{DBLP:conf/tacas/MouraB08}. However, it leads to a further problem which is lack of diversity due to that existing constraint solvers tend to pick boundary values for constraints.
To relieve this problem, NNSmith tries to iteratively add extra constraints (named ``attribute binning''~\cite{NNSmith}). When extra constraints produce an unsatisfiable one, NNSmith will randomly drop some of the constraints and retry, until it succeeds.
It results in following disadvantages: (1) extra overhead for constraint solving due to the iteratively retrying; (2) a biased distribution of operations and parameters in the generated models, the models tend to contain more ``simple'' operations such as \CodeIn{Add} and \CodeIn{Sub}, and less ``complicated'' operations such as \CodeIn{Conv} and \CodeIn{Gemm} (as seen in both of the results in NNSmith's paper~\cite{NNSmith} and our evaluation).

Compared with related work, our test generation approach overcomes the validity challenge by the domain-specific constraint solver proposed in this paper. It offers several advantages as follows: 
\begin{itemize}
    \item The computation graphs generated by our approach are more diverse because our domain-specific constraint solver can sample diverse operations/parameters without bias, as evidenced in our evaluation.
    \item
    Our approach is more efficient due to lower computational costs compared to other solutions such as repeatedly calling the external constraint solver (as NNSmith did), as evidenced in our evaluation.
    \item Our approach is more scalable because our domain-specific constraint solver is lightweight and backtrack-free,
    without inherent limitations on the type and the size of generated computation graphs, which is potentially beneficial for other scenarios such as generating extreme test cases for stress testing.
\end{itemize}

Besides test generation for valid computation graphs, existing research work also proposes other techniques to enhance the effectiveness of deep learning compiler testing, which are orthometric to our approach.
NNSmith~\cite{NNSmith} conducts value searching for improving numeric validity with gradients. HirGen~\cite{HirGen} proposes ``disruptive generation'' to generate computation graphs containing obvious breaks of specifications for detecting incorrect exception handling bugs.
In addition, mutation-based approaches such as  TVMFuzz~\cite{DBLP:conf/sigsoft/ShenM0TCC21} and Tzer~\cite{DBLP:journals/pacmpl/LiuWYDZ22}, 
conduct a series of heuristic-based mutation rules on seed inputs (i.e., existing models) at the compiler's low-level intermediate representation.
Our approach is complementary to these techniques.
}

\section{Conclusion}
In this paper, to construct diverse and semantically valid computation graphs for testing deep learning compilers, we proposed a new approach named \toolname{}, including a novel domain-specific solver for effectively resolving constraints on computation graphs.
We have implemented and evaluated our approach against \pre{four} \rev{five} baselines, and also applied \toolname{} to test three real-world deep learning compilers. The evaluation results show that 
\pre{
(1) \toolname{} outperforms the baselines including two state-of-art approaches (Muffin~\cite{DBLP:conf/icse/GuLZ022} and TVMFuzz~\cite{DBLP:conf/sigsoft/ShenM0TCC21})
on both coverage metrics and bug-detection capability; (2) \toolname{} detects critical bugs in the released versions of the three compilers, demonstrating its high value in practice.
}
\rev{
(1) \toolname{} outperforms the baselines including two state-of-the-art approaches (Muffin~\cite{DBLP:conf/icse/GuLZ022} and NNSmith~\cite{NNSmith})
on coverage metrics, demonstrating \toolname{}'s effectiveness in generating diverse computation graphs;
(2) \toolname{} performs better or as well than state-of-the-art approaches on bug detection, the result of \toolname{} is also complementary to those from existing approaches; (3) \toolname{} detects 24 previously unknown bugs in the released versions of the three compilers, demonstrating its high value in practice.
}

\section{Acknowledgments}
This work was partially supported by National Natural Science Foundation of China under Grant No. 62161146003, and the Tencent Foundation/XPLORER PRIZE.

We would like to thank Haiyue Ma, Zhengkai Wu, Chenxi Li for their help with improving the presentation of this work.

\clearpage

\bibliographystyle{ieee_fullname}
\bibliography{sample-base}

\end{document}